\RequirePackage{amsmath}
\documentclass[a4paper]{llncs}
\usepackage{hyperref,graphicx,setspace,enumitem, booktabs,multirow,tikz,doi,color,colortbl,algorithm,algorithmicx,cleveref,csquotes,subcaption}
\usepackage[noend]{algpseudocode}
\usepackage[margin=1in]{geometry}

\usepackage{etoolbox}
\apptocmd{\sloppy}{\hbadness 10000\relax}{}{}

\newcommand{\vs}{\vspace{0.2cm}}
\newcommand{\clrone}{\rowcolor{lightgray!50}}
\newcommand{\clrtwo}{\rowcolor{lightgray}}

\newcommand{\bottomline}{\noalign{\global\arrayrulewidth=0.3mm}\arrayrulecolor{lightgray}\hline}
\newcommand{\stef}[1]{\textcolor{red}{Stef: {#1}}}
\renewcommand{\b}{\bfseries}
\newcommand{\mb}[1]{\mathbf{#1}}
\renewcommand{\(}{\left(}
\renewcommand{\)}{\right)}
\newcommand{\ph}{$\phantom{^{\ast}}$}
\newcommand{\tictoc}[1]{\textcolor{magenta}{Con: {#1}}}
\newcommand{\qt}[1]{\enquote{#1}}

\begin{document}

\title{A Peek into the Unobservable: Hidden States and Bayesian Inference for the Bitcoin and Ether Price Series}
\author{Constandina Koki\inst{1} \and
Stefanos Leonardos\inst{2} \and
Georgios Piliouras\inst{2}}
\institute{Athens University of Economics and Business, 76 Patission Str. GR-10434 Athens, Greece \email{\{kokiconst,vrontos\}@aueb.gr} \and
Singapore University of Technology and Design, 8 Somapah Rd. 487372 Singapore, Singapore \email{\{stefanos\_leonardos,georgios\}@sutd.edu.sg}
}

\maketitle        

\begin{abstract}
Conventional financial models fail to explain the economic and monetary properties of cryptocurrencies due to the latter's dual nature: their usage as financial assets on the one side and their tight connection to the underlying blockchain structure on the other. In an effort to examine both components via a unified approach, we apply a recently developed Non-Homogeneous Hidden Markov (NHHM) model with an extended set of financial and blockchain specific covariates on the Bitcoin (BTC) and Ether (ETH) price data. Based on the observable series, the NHHM model offers a novel perspective on the underlying microstructure of the cryptocurrency market and provides insight on unobservable parameters such as the behavior of investors, traders and miners. The algorithm identifies two alternating periods (hidden states) of inherently different activity -- fundamental versus uninformed or noise traders -- in the Bitcoin ecosystem and unveils differences in both the short/long run dynamics and in the financial characteristics of the two states, such as significant explanatory variables, extreme events and varying series autocorrelation. In a somewhat unexpected result, the Bitcoin and Ether markets are found to be influenced by markedly distinct indicators despite their perceived correlation. The current approach backs earlier findings that cryptocurrencies are unlike any conventional financial asset and makes a first step towards understanding cryptocurrency markets via a more comprehensive lens. 
\end{abstract}

%\section{Highlights}
%\stef{Here we should phrase in 4-5 bullets the main points of the paper, the things that we want to be mentioned by the people (if any) who will cite the paper. Like all papers in sciencedirect.}
%
%\begin{itemize}
%\item We use a NHHM (regime switching) model, to study the properties of the  BTC and ETH series using financial and cryptograpic specific covariates.
%%\item We use financial/economic and cryptocurrency specific covariates to study the semi-strong efficiency of the cryptocurrencies conditional on the NHPG model.
%\item The studied model gives insides on the log-retruns BTC series in contrast with the log-returns ETH series which can not be explained under this Hidden Markov setting.
%\item Using financial/economic covariates the forecasting accuracy is poor.
%\item The NHPG model identifies two subchains of the BTC series with different behavior/characteristics, such as volatility, kurtosis, skewness, stationarity, that can be useful to investors.
%\item BTC is more efficient than ETH (efficiency increases with age). It is more correlated to global indicators than ETH.
%\end{itemize}

\section{Introduction}
\subsection{Motivation, Methodology and Main Results}
The present study is motivated by the still limited understanding of the economic and financial properties of cryptocurrencies. Sheding light on such properties constitutes a necessary step for their wider public adoption and is fundamental for blockchain stakeholders, investors, interested authorities and regulators (\cite{Co19,Li19,Pre20}). More importantly, it may provide hints about market manipulation and fraud detection. \par
Unfortunately, existing financial models that are used to study fiat currency exchange rates fail to capture the convoluted nature of cryptocurrencies (\cite{Ca19}). The additional challenge that they face is the tight connection between cryptocurrency prices and the underlying blockchain technology which drives the dynamics of the observable market. To some extent, this is expressed via the particular market microstructure of cryptocurrencies: the market depth which depends on the exchange and the market maker, the functionality of exchanges as custodians (unique property among financial assets) and the absence of stocks, equities or other financial investment instruments (with the exception of Bitcoin futures, \cite{Ka19}) which render acquiring and/or trading the cryptocurrency the main way of investing in this new technology (\cite{Ko18}). The miners and/or stakers emerge as the main actors who drive the creation and distribution of the currency whereas the cheap and immediate transactions essentially obviate the need for conventional brokers. All these features (among many others), starkly distinguish cryptocurrencies from conventional financial assets or fiat money. However, a precise understanding of their defining financial and economic properties is still elusive (\cite{Br18,Co18,Ur18}). With this in mind, the concrete research questions that we set out to understand are the following:  
\begin{itemize}[leftmargin=*]
\item How do cryptocurrencies compare -- in terms of their economic and financial properties -- to well understood financial assets like commodities, precious metals, equities and fiat currencies (\cite{Li17,Ba18,Kan19})? How do they relate to traditional financial markets and global macroeconomic indicators? 
\item What are the defining microstructure characteristics of the cryptocurrency market and which are the distinguishing features (if any) between different coins (\cite{Kat19})?
%\item What drives cryptocurrency prices? Despite their evident connectedness, , do they show signs of differentiation and if yes, what are the variables that govern these differences? 
\end{itemize}
To address these questions, we use a recently developed instance of Non-Homoge\-neous Hidden Markov (NHHM) modeling, namely the Non-Homogeneous P\'olya Gamma Hidden Markov model (NHPG) of \cite{Ko19,Kok20b}, which has been shown to outperform similar models in conventional financial data (\cite{Me11}). Using financial and blockchain specific covariates on the Bitcoin (\cite{Na08}) and Ether (\cite{Bu14,But20}) log-return series (henceforth BTC and ETH, respectively), the NHHM methodology aims not only to capture dynamic patterns and statistical properties of the observable data but more importantly, to shed some light on the unobservable financial characteristics of the series, such as the activity of investors, traders and miners. \par
The present model falls into the Markov-switching or regime-switching literature with two possible states that is the benchmark for predicting exchange rates (\cite{En94,Le06,Fr05,Be16,Gr13,Wr09}). % It uses Bayesian Model Averaging (BMA) for inference which has been shown to possess desirable properties for forecasting applications (\cite{Be16,Gr13,Wr09} and \cite{Wr08}). 
This linear model was first introduced by \cite{Ha89} as an alternative approach to model non-linear and non-stationary data. It involves switches between multiple structures (equations) that can characterize the time series behavior in different regimes (states). The switching mechanism is governed by an unobservable state variable that follows a first-order Markov chain\footnote{For example, in the seminal paper of \cite{Ha89}, the author used the underlying hidden process to define the business cycles (recession periods). More recent examples and a comprehensive theory about NHHM in finance can be found in \cite{Ma14}.}. Therefore, the NHMM is suitable for describing correlated and heteroskedastic data with distinct dynamic patterns during different time periods, as are precisely cryptocurrency prices (\cite{Agg19,Bo19a,Kat19}). \par
Although standard in financial applications (\cite{Ma14}), Hidden Markov models have only been applied in the cryptocurrency context by \cite{Po19} as state space models, by \cite{Ko18} to capture the liquitity uncertainty and \cite{Ph17} in the context of price bubbles. Yet, their more extensive use is supported by the specific characteristics of cryptocurrency data that have been identified by earlier research. \cite{Ba17,Ja18} and \cite{De18} demonstrate the non-stationarity of BTC prices and volume and underline the importance of modeling non-linearities in Bitcoin prediction models. This is further elaborated by \cite{Be16,Pi17,Ph18} and \cite{Yu19} who suggest that model selection and the use of averaging criteria are necessary to avoid poor forecasting results in view of the cryptocurrencies' extreme and non-constant volatility. Along these lines, \cite{Ci16} show that the Bitcoin price series exhibits structural breaks and suggest that significant price predictors may vary over time. Additional motivation for the analysis of cryptocurrency data with regime-switching models as the one employed here, is provided by \cite{Ka17} who demonstrate the heteroskedasticity of BTC prices and \cite{Bau18} who identify periods of different trading activity. Our main findings can be summarized as follows 
\begin{itemize}[leftmargin=*]
\item The NHPG algorithm identifies two hidden states with frequent alternations for the BTC log-return series, cf. \Cref{fig:plotbtc}. State 1 corresponds to periods with higher volatility and returns and accounts for roughly one third of the sample period (2014-2019). By contrast, state 2 marks periods with lower volatility, series autocorrelation (long memory), trend stationarity and random walk properties, cf. \Cref{tab:subchains}. At the more variable state 1, the BTC data series is influenced by miners' activity and more volatile covariates (stock indices) in comparison to more stable indicators (exchange rates) in state 2, cf. \Cref{tab:postbtc}. 
\item The results for the hidden process are the same for both the long run (2014-2019) and the short run (2017-2019) BTC data, cf. \Cref{fig:plotbtc,fig:btc_small}. However, differences in the significant predictors indicate more speculative activity in the short run compared to more fundamental investor behavior in the long run, cf. \Cref{tab:postbtc}. In sum, speculative activity (noise traders) is identified in the less frequent state 1 and in the short run whereas increased activity of fundamental investors is seen in state 2 and in the long run.
\item The algorithm does not mark a well defined hidden process with clear transitions for the ETH series, cf. \Cref{fig:ploteth}. This is further supported by the low number and the small values of significant predictors from the current set, cf. \Cref{tab:posteth}. These results imply that ETH prices are still driven by variables beyond the currenlty selected set of predictors, showing characteristics of an emerging market that is more isolated than BTC from global financial and macroeconomic indicators.  
\end{itemize}
More details are presented in \Cref{sec:results}. Overall, the outcome of the NHPG model can be useful for investors and blockchain stakeholders by providing hints on periods of differentiating activities and effects in the cryptocurrency markets. From a theoretical perspective, it backs earlier findings that cryptocurrencies are unlike any other financial asset and suggests that their understanding requires not only the integration of existing financial tools but also a more refined framework to account for their bundled technological and financial features (\cite{Co18}).

\subsection{Related Literature}
The literature on the financial properties of cryptocurrencies is expanding at an exponential rate and an exhaustive review is not possible (see \cite{Cor19} and references therein for a more comprehensive reference list). More relevant to the current context is the scarcity (to the best of our knowledge) of papers that address the bundled nature of cryptocurrencies as both blockchain applications and financial assets. Existing studies focus either on the underlying blockchain technology/consensus mechanism or on the observable financial market but not on both. By contrast, the current NHPG model parses the observable financial information to recover the underlying structure of cryptocurrency markets and hence makes a first step towards a unified approach to fill this gap. Its limitations are discussed in \Cref{sec:discussion}. In the remaining part of this section, we provide a (non-exhaustive) list of studies that focus on the financial part. \par
Early research, mainly focusing on BTC has provided mixed insights on the properties of cryptocurrencies. \cite{Kl18} claim that BTC is fundamentally different from valuable metals like gold due to its shortage in stable hedging capabilities. Along with \cite{Ch15}, \cite{Ci16} also argue that standard economic theories cannot explain BTC price formation and using data up to 2015, they provide evidence that BTC lacks the necessary qualities to be qualified as money. However, \cite{Dy16} demonstrate that BTC has similarities to both gold and the US dollar (USD) and somewhat surprisingly, that it may be ideal for risk-averse investors. \cite{Bo17,Bo17a} and \cite{Bo17b} also explore BTC's characteristics as a financial asset and find that while BTC is useful to diversify financial portfolios -- due to its negative correlation to the US implied volatility index (VIX) -- it otherwise has limited \emph{safe haven} properties. Using data from a longer period (between 2010 and 2017), \cite{De18} conclude the opposite, namely that BTC may indeed serve as a hedging tool, due to its relationship to the Economic Policy Uncertainty Index (EUI). In comparative studies, \cite{Fr18,Corb18} provide empirical evidence of bubbles in both BTC and ETH and \cite{Gk18} suggest that BTC is less risky than ETH, i.e., that it exhibits less fat tailed behavior. \cite{Phi18} confirm that Bitcoin exhibits long memory and heteroskedasticity and argue that cryptocurrencies display mild leverage effects, predictable patterns with mostly oscillating persistence, varied kurtosis and volatility clustering. Comparing BTC with ETH, they argue that kurtosis is lower for ETH being easier to transact than BTC. Along this line, the findings of \cite{Me19} and \cite{Kat19} further motivate the use of non-homogeneous and regime-switching modeling for both the BTC and ETH log-returns series.\par
The differences between cryptocurrencies and conventional financial markets are further elaborated by \cite{Ka17,Ha17,Ph18}. High volatility, speculative forces and large dependence on social sentiment at least during its earlier stages are shown by some as the main determinants of BTC prices (\cite{Ga15,Ge15,Co15,Yi18}). Yet, a large amount of price variability remains unaccounted for (\cite{Ho18,Mc18,Ja18}). Moreover, the proliferation of cryptocurrencies on different blockchain technologies suggests that their current correlation may be discontinued in the near future and calls for comparative studies as the one conducted here (\cite{Bo19}).

\subsection{Outline}
The rest of the paper is structured as follows. In \Cref{sec:methodology}, we describe the NHPG model and simulation scheme and present the set of variables that have been used (some preliminary descriptive statistics and tests about this data are relegated to \Cref{app:appendix}). \Cref{sec:results} contains the main results and their analysis. In the first part (\Cref{sec:hidden_btc,sec:hidden_btceth,sec:comparisons}), we present the outcome of the algorithm and discuss the statistical findings for the hidden states and the generated subseries. In the second part, \Cref{sec:covariates}, we focus on the significant explanatory variables for the BTC data series in both the short and long run and the ETH data series. We conclude the paper with a discussion of the limitations of the present model and directions for future work in \Cref{sec:discussion}. 
 
\section{Methodology \& Data}\label{Methodology}\label{sec:methodology}
Given a time horizon $T\ge0$ and discrete observation times $t=1,2,\dots,T$, we consider an observed random process $\left\{Y_{t}\right\}_{t\le T}$ and a hidden underlying process $\left\{Z_{t}\right\}_{t\le T}$. The hidden process $\left\{Z_t\right\}$ is assumed to be a two-state non-homogeneous discrete-time Markov chain that determines the states ($s$) of the observed process. In our setting, the observed process is either the BTC or the ETH log-return series. Importantly, the description of the hidden states is not pre-determined and is subject to the outcome of the algorithm and interpretation of the results.\par
Let $y_{t}$ and $z_{t}$ be the realizations of the random processes $\left\{Y_{t}\right\}$ and $\{Z_t\}$, respectively. We assume that at time $t,\ t=1,\dots,T$, $y_{t}$ depends on the current state $z_{t}$ and not on the previous states. Consider also a set of $r-1$ available predictors $\left\{X_{t}\right\}$ with realization $x_{t}=(1,x_{1t},\dots ,x_{r-1t})$ at time $t$. The explanatory variables (covariates) $\left\{X_t\right\}$ that are used in the present analysis are described in \Cref{tab:variables}. The effect of the covariates on the cryptocurrency price series $\{Y_t\}$ is twofold: first, linear, on the mean equation and second non-linear, on the dynamics of the time-varying transition probabilities, i.e., the probabilities of moving from hidden state $s=1$ to the hidden state $s=2$ and vice versa. Given the above, the cryptocurrency price series $\{Y_{t}\}$ can be modeled as 
\[Y_t\mid Z_t=s \sim \mathcal{N}(x_{t-1}B_s,\sigma^2_{s}), \;s=1,2,\]
where $B_{s}=(b_{0s},b_{1s},\dots ,b_{r-1 s})'$ are the regression coefficients and $\mathcal{N}(\mu,\sigma^2)$ denotes the normal distribution with mean $\mu$ and variance $\sigma^2$. The dynamics of the unobserved process $\left\{Z_{t}\right\}$ can be described by the time-varying (non-homogeneous) transition probabilities, which depend on the predictors and are given by the following relationship 
$$P(Z_{t+1}=j\mid Z_t=i)=p^{(t)}_{ij}=\frac{\exp(x_{t}\beta_{ij})}{\sum^{2}_{j=1}\exp(x_t\beta_{ij})}, \; i,j=1,2,$$ 
where $\beta _{ij}=(\beta _{0,ij},\beta _{1,ij},\dots ,\beta_{r-1,ij})^{\prime }$ is the vector of the logistic regression coefficients to be estimated. Note that for identifiability reasons, we adopt the convention of setting, for each row
of the transition matrix, one of the $\beta_{ij}$ to be a vector of zeros. Without loss of generality, we set $\beta_{ij}=\beta_{ji}=\mathbf{0}$ for $i,j=1,2, i\neq j$. Hence, for $\beta_i:=\beta_{ii},\;i=1,2$, the probabilities can be written in a simpler form $$p^{(t)}_{ii}=\frac{\exp(x_{t}\beta_{i})}{1+\exp(x_t\beta_{i})} \ \text{and}\   p^{(t)}_{ij}=1-p^{(t)}_{ii} ,\ i,j=1,2,\ i\neq j.$$ 
To make inference on the hidden process, we use the smoothed marginal probabilities $P(Z_t=i\mid Y_{1:T},z_{t+1},\theta)$ which are the probabilities of the hidden state conditional on the full observed process as derived from the Forward-Backward algorithm (\cite{Ha89}). In the rest of the paper, we use the notation $P\(Z_t=i\)$ for convenience.

\begin{algorithm}[!htb]
\caption{MCMC Sampling Scheme for Inference on Model Specification and Parameters}\label{alg:mcmc}
\begin{algorithmic}[1]
\State{\textit{\% After each procedure the parameters and model space are updated conditionally on the previous quantities}}\vspace{0.1cm}
\Procedure{Scaled Forward Backward}{$\(z_{1:T}\)$}
\State{\textit{\%Simulation of a realization of the hidden states $z_t$}}
\For{$t=1,\dots,T$ and $i=1,2$}
\State {$\pi_t\(i\mid\theta\)\gets \dfrac{\alpha_{t}(s)}{\sum^{2}_{j=1}\alpha_t\(j\)}=P\(z_t=i\mid\theta,y^{t}\)$} \\
$\Comment{\text{Simulation of the scaled forward variables}}$
\EndFor
\For {$t=T,T-1,\dots,1$}
\State $z_t \gets P\(z_t\mid y^T, z_{t+1}\)=\dfrac{p_{iz_{t+1}}\pi_t\(i\mid\theta\)}{\sum_{j=1}^{m}p_{jz_{t+1}}\pi_t\(j\mid\theta\)}$\\ $\Comment{\text{Backwards simulation of $z_t$ using the smoothed probabilities}}$\vspace{0.1cm}
\EndFor
\EndProcedure
\Procedure{Mean\_Regres\_Param}{$B_s, \sigma^2_s, s=1,2$}
\State{\textit{\%Simulation of the mean regression parameters}}
\For{$s=1,2$} $\Comment{\text{Conjugate analysis with Gibbs sampler}}$
\State {$B\mid\sigma^2 \sim f_B, \sigma^2 \sim \mathcal{IG}$} $\Comment{\text{$f_B\equiv$ Normal and $\mathcal{IG} \equiv$ Inverse-Gamma}}$\vspace{0.1cm}
\EndFor
\EndProcedure\vspace{0.1cm}
\Procedure{Log\_Regres\_Coef}{$\(\beta_s, \omega_s \)$}
\State{\textit{\%Simulation of the logistic regression coefficients}}
\For{$s=1,2$} $\Comment{\text{P\'{o}lya-Gamma data augmentation scheme}}$
\State {$\to$ augment the model space with $\omega_s$} \\$\Comment{\text{Conjugate analysis on the augmented space}}$
\State{$\to$ sample from $\beta_s\sim f_{\beta_s\mid \omega}$ and $\omega_s\mid \beta_s\sim \mathcal{PG}$}
\State{$\to$ posteriors $f_{\beta_s\mid \omega}\equiv$ Normal and $\mathcal{PG}\equiv$ P\'{o}lya-Gamma}\vs
\EndFor
\EndProcedure
\end{algorithmic}\vspace{-0.2cm}
\end{algorithm}

\subsection{Simulation Scheme}

The unknown quantities of the NHPG are $\left\{\theta_s=\(B_{s},\sigma _{s}^{2}\),\beta_s, s=1,2 \right\}$, i.e., the parameters in the mean predictive regression equation and the parameters in the logistic regression equation for the transition probabilities. We follow the methodology of \cite{Ko19}. In brief, the authors propose the following MCMC sampling scheme for joint inference on model specification and model parameters.
\begin{enumerate}[itemsep=0cm]
\item Given the model's parameters, the hidden states are simulated using the Scaled Forward-Backward of algorithm of \cite{Sc02}.
\item The posterior mean regression parameters are simulated using the standard conjugate analysis, via a Gibbs sampler method.
\item The logistic regression coefficients are simulated using the P\'{o}lya-Gamma data augmentation scheme \cite{Po13}, as a better and more accurate sampling methodology compared to the existing schemes.
\end{enumerate} 
The steps 1-3 of the MCMC algorithm are detailed in \Cref{alg:mcmc}.

\subsection{Data}\label{sec:data}
\begin{figure}[!htb]
\centering
\includegraphics[width=\linewidth, trim=0cm 2cm 0cm 1cm]{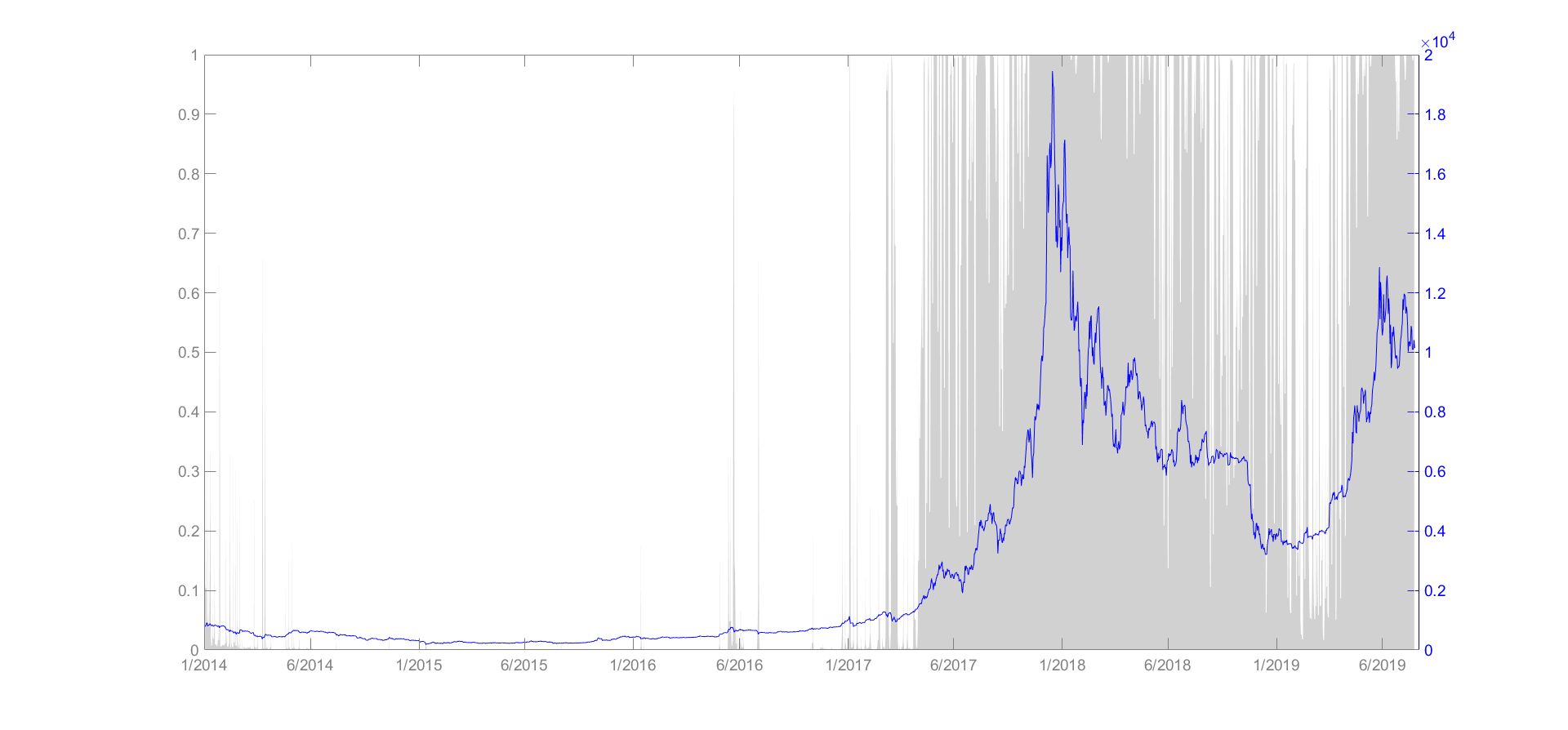}
\caption{Application of the NHPG model on the BTC price series. The algorithm essentially identifies two periods, the first from 2014 (start of the dataset) to 2017 and the second from 2017 to date. This motivates separate analysis of the BTC for the latter period and comparison with the ETH price series over the same period.}\vspace{-0.5cm}
\label{fig:pricebtc}
\end{figure}

We assess the ability of 11 financial--macroeconomic and 3 cryptocurrency specific variables, outlined in \Cref{tab:variables}, in explaining and forecasting the prices of BTC and ETH via the NHPG model. In the related cryptocurrency literature these indices are commonly studied under various settings 
(\cite{Wi13,Ye15,Bo17b,Es17,Pi17,Ho18}, \cite{Ja18} and \cite{Po19}). The findings of the descriptive statistics and preliminary stationarity tests, cf. \Cref{app:appendix}, indicate that the logarithmic return (log-return), i.e., the change in log price, $r_t=\log\(y_t\)-\log\(y_{t-1}\)$, series of BTC and ETH exhibit trend non-stationarity, non-linearities, rich (i.e., non-random) underlying information structure and non-normalities. Based on these properties, the NHPG model seems appropriate for the study of the log-return data series. Accordingly, we apply the NHPG algorithm on daily log-returns of BTC and ETH, with normalized explanatory variables. We perform two experiments over two different time frames: in the first, we study the BTC series between 1/2014 and 8/2019 and in the second, we study both the BTC and ETH series between 1/2017 and 8/2019. The second time frame has been selected to allow reasonable comparisons between the BTC and ETH prices after eliminating an initial period following the launch of the ETH currency. It is further motivated by the outcome of a test-run of the NHPG model on BTC prices, cf. \Cref{fig:pricebtc}, which indicates a transition point to a different period for the BTC price series in January 2017. 

\begin{table}[!htb]
\centering
\setlength{\tabcolsep}{10pt}
\renewcommand{\arraystretch}{1.3}
%\resizebox{\textwidth}{!}{%
\begin{tabular}{lll}
\clrtwo
\multicolumn{3}{c}{Explanatory Variables}\\ 
Description & Symbol & Retrieved from\\[0.02cm]
\bottomline\\[-0.4cm]
\clrone US dollars to Euros exchange rate & USD/EUR & \href{https://www.investing.com/}{investing.com}\\
\clrone US dollars to GBP exchange rate & USD/GBP & \href{https://www.investing.com/}{investing.com}\\
\clrone US dollars to Japanese Yen exchange rate & USD/JPY & \href{https://www.investing.com/}{investing.com}\\
\clrone US dollars to Chinese Yuan exchange rate & USD/CNY &\href{https://www.investing.com/}{investing.com}\\\\[-0.4cm]
Standard \& Poor's 500 index & SP500 &\href{https://finance.yahoo.com/}{finance.yahoo.com}\\
NASDAQ Composite index  & NASDAQ &\href{https://finance.yahoo.com/}{finance.yahoo.com}\\[0.1cm]
\clrone Silver Futures price&Silver&\href{https://www.investing.com/}{investing.com}\\
\clrone Gold Futures price & Gold &\href{https://www.investing.com/}{investing.com}\\
\clrone Crude Oil Futures price & Oil &\href{https://www.investing.com/}{investing.com}\\[0.1cm]
CBOE Volatility index & VIX &\href{https://finance.yahoo.com/}{finance.yahoo.com}\\
Equity market related Economic Uncertainty index & EUI &\href{https://fred.stlouisfed.org}{fred.stlouisfed.org}\\[0.1cm]
\clrone Daily Block counts& Blocks &\href{https://www.coinmetrics.io}{coinmetrics.io}\\
\clrone Hash Rate & Hash &\href{https://www.quandl.com/}{quandl.com}, \href{https://etherscan.io}{etherscan.io}\\
\clrone Transfers of native units & Tx-Units &\href{https://www.coinmetrics.io/}{coinmetrics.io}\\ \\[-0.4cm]	
\bottomline
\end{tabular}
%}
\vs
\caption{List of variables and online resources. The Hash Rate (Hash) has been retrieved from \href{https://www.quandl.com/}{quandl.com} for Bitcoin (BTC) and from \href{https://etherscan.io}{etherscan.io} for Ether (ETH).}
\label{tab:variables}\vspace{-0.3cm}
\end{table}

\section{Results \& Analysis}\label{sec:results}
In this section, we discuss the findings from the NHPG model on the BTC and ETH log-return series. We first present the graphics with the output of the algorithm for the whole 2014-2019 period on BTC log-returns (\Cref{sec:hidden_btc}) and the shorter 2017-2019 period on both BTC and ETH log-returns (\Cref{sec:hidden_btceth}). Then, we interpret the results and compare the statistical properties and the significant covariates between the two hidden states of both the BTC and ETH series and between the short and long run BTC series (\Cref{sec:covariates,sec:comparisons}).

\subsection{Hidden States: Bitcoin 2014--2019}\label{sec:hidden_btc}
\begin{figure}[!htb]
\centering
\includegraphics[width=\linewidth, trim=2cm 1cm 2cm 3cm]{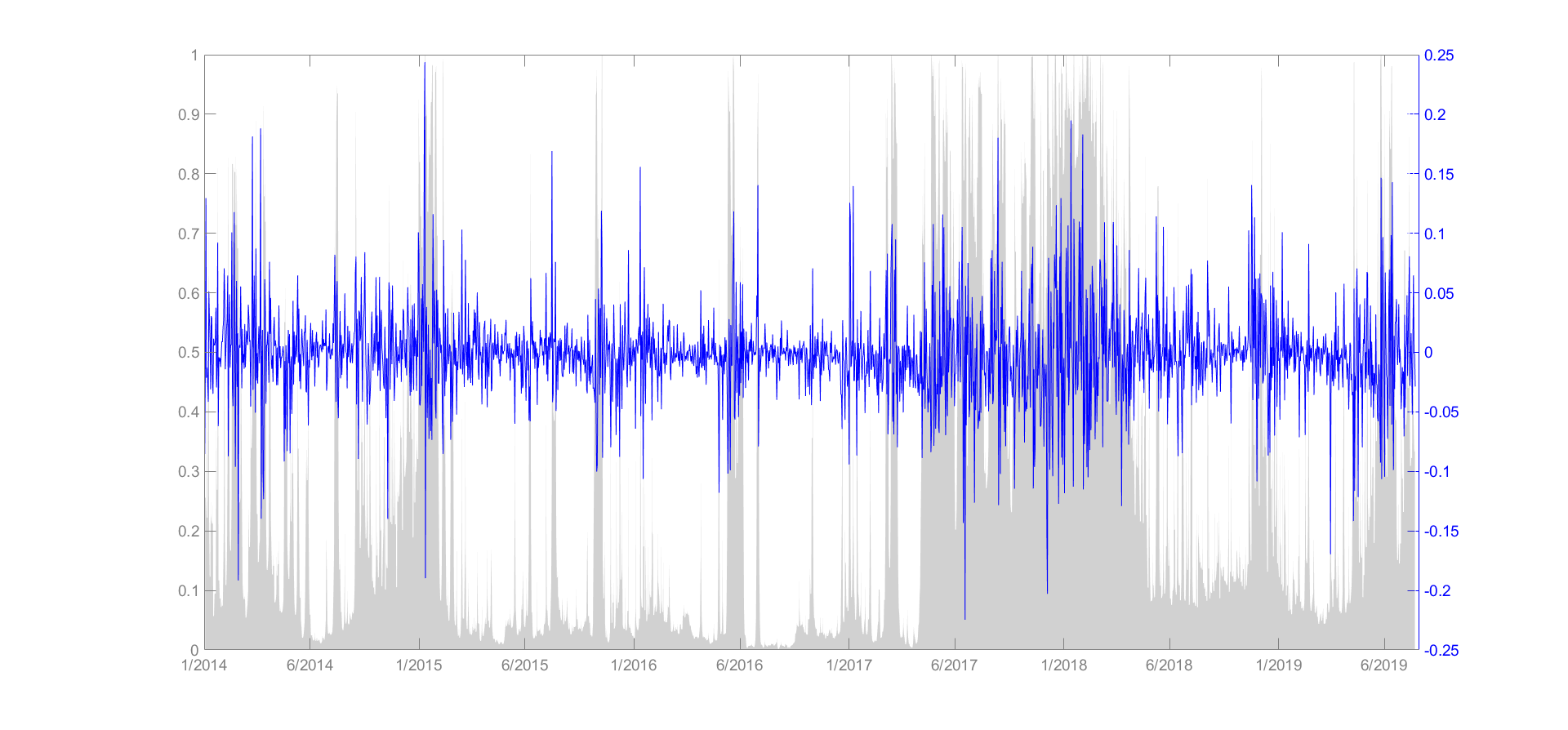}
\caption{BTC logarithmic-return series (blue line -- right axis) for the period 1/2014-8/2019 with the mean smoothed marginal probabilities of state 1, i.e., $Pr\(Z_t=1\)$ (gray bars -- left axis).}
\label{fig:plotbtc}
\end{figure}

\Cref{fig:plotbtc} displays the BTC log-return series (blue line) along with the smoothed marginal probabilities (gray bars) of the hidden process being at state 1. Using as a threshold the probability $P\(Z_t=1\)>0.5$, we estimate the hidden states for each time period. The NHPG model identifies a subseries of 667 observations in state 1 and a subseries of 1388 observations in state 2. The description of the hidden states is not predetermined by the model and is done a posteriori, by comparison of the statistical properties of the two subseries that have been generated. As it is obvious from \Cref{fig:plotbtc}, state 1 corresponds to periods of larger log-returns and increased volatility in comparison to state 2. The frequent changes are in line with previous studies on the heteroskedasticity and on the regime switches (structural breaks) of the Bitcoin time series (\cite{Ph17,Ka17} and \cite{Ko18,Ar19}, respectively). Yet, the refined outcome of the NHPG model, which determines the time periods that the series spends in each state, allows for a more granular approach. Specifically, it adds information about the significant covariates that affect both the observable and the unobservable process and on the financial properties of each state. This is done in \Cref{sec:covariates} below.

\subsection{Hidden States: Bitcoin and Ether 2017--2019}\label{sec:hidden_btceth}
\begin{figure}[!b]
\centering
\hspace{-0.5cm}
\begin{subfigure}[b]{0.5\textwidth}
\includegraphics[width=\textwidth, trim=7cm 1cm 0cm 2cm]{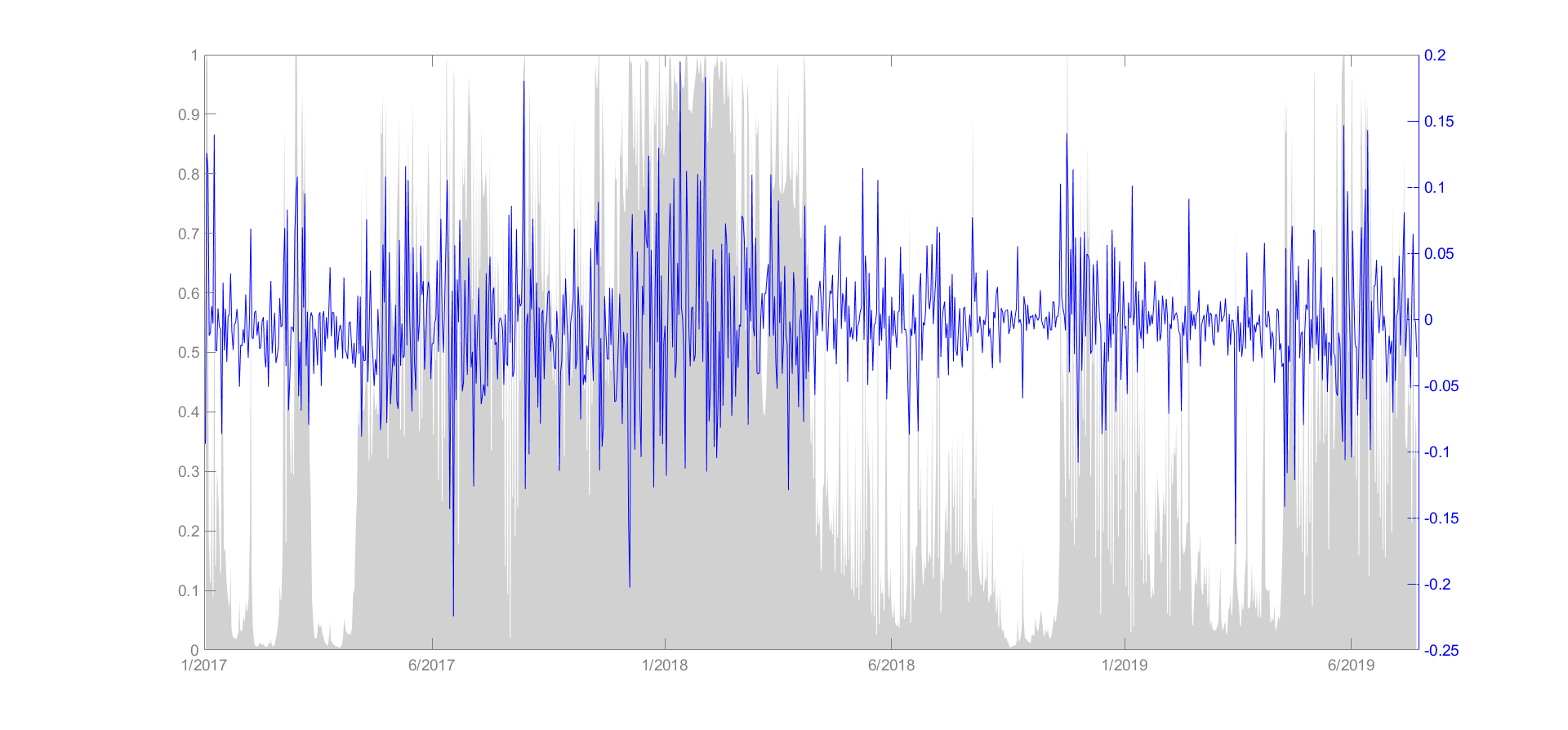}
\caption{BTC: 1/2017-8/2019}
\label{fig:btc_small}
\end{subfigure}\hspace{-17pt}
\begin{subfigure}[b]{0.5\textwidth}
\includegraphics[width=\textwidth, trim=0cm 1cm 7cm 2cm]{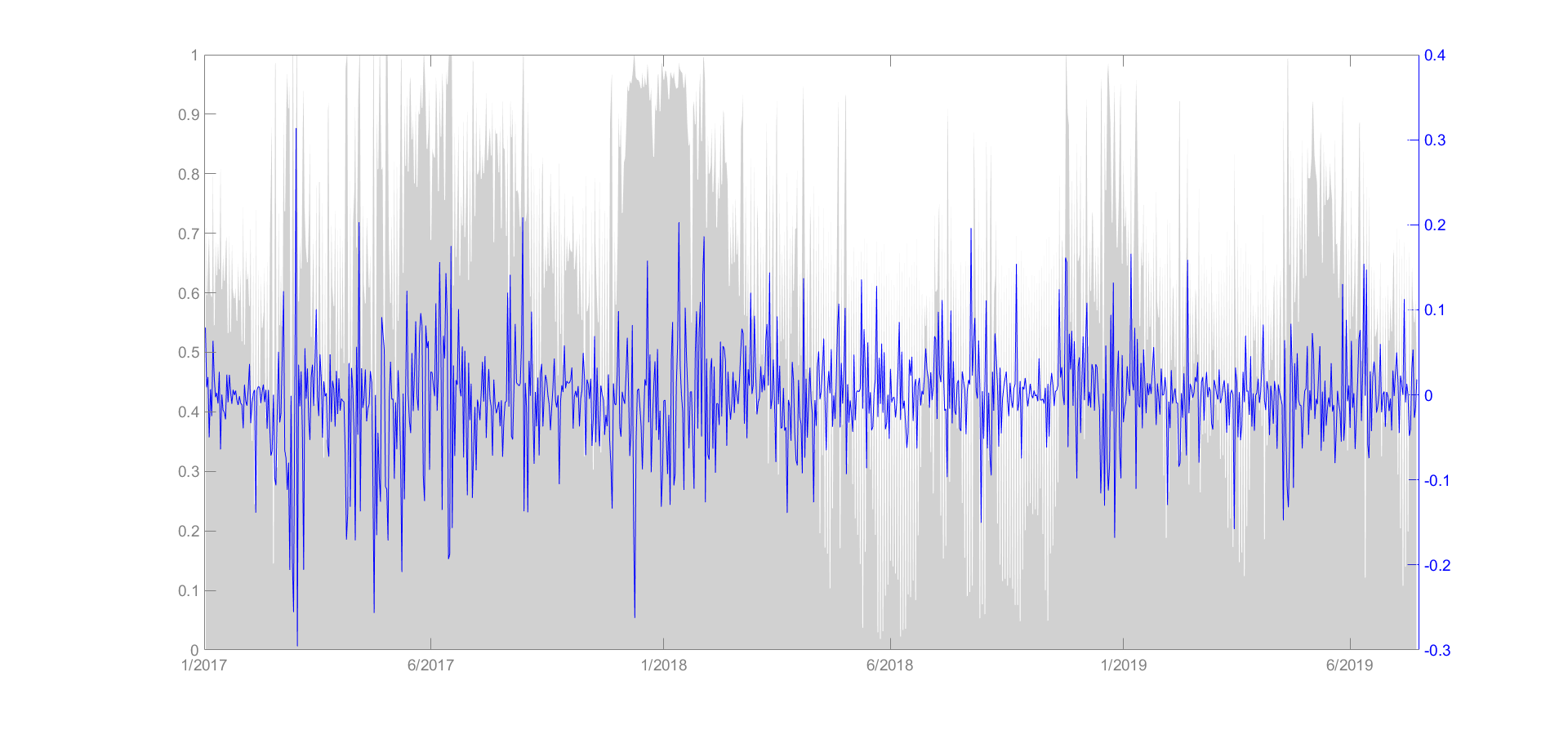}
\caption{ETH: 1/2017-8/2019}
\label{fig:ploteth}
\end{subfigure}
\vskip0.3cm
\caption{BTC (left panel) and ETH (right panel) logarithmic-returns series (blue lines -- right axis) for the period 1/2017-8/2019 with the mean smoothed marginal probabilities of state 1, i.e., $Pr\(Z_t=1\)$, (gray bars -- left axis).}
\label{fig:compare}\vspace{-0.5cm}
\end{figure}

\Cref{fig:compare} shows the results of the NHPG model for both the BTC (left panel) and ETH (right panel) log-return series over the shorter 1/2017-8/2019 period. The algorithm has again identified two states in the BTC series, \Cref{fig:btc_small}, as indicated by the clear distinction between high-low marginal probabilities of state 1, i.e., $P\(Z_t=1\)$, that are given by the gray bars. Moreover, a comparison with the same period in \Cref{fig:plotbtc} demonstrates that the NHPG has produced the same result (zoom in) -- in terms of statistical quality -- even over this smaller period, i.e., the algorithm has converged and returns essentially the same probabilities for the underlying process. However, as we will see below, cf. \Cref{sec:covariates}, the statistical analysis unveils differences in the significant predictors and financial properties between the short and long run.

The picture is different for the ETH series, cf. \Cref{fig:ploteth}. Here, the hidden process is not well defined since the probabilities of state 1 at each time period are mostly close to $0.5$. This indicates high degree of randomness in the transitions of the algorithm and along with the low number of significant covariates that have been identified for ETH (cf. \Cref{tab:posteth} below), it suggests that ETH prices are still influenced by forces which are beyond the current set of financial and blockchain indicators (\cite{Ka18,Ph18,Phi18}). This implies that ETH -- when viewed as a financial asset -- shows characteristics of an evolving, non-static and still emerging market. However, the relative isolation of ETH from other financial assets agrees with earlier findings in the literature (\cite{Ph18,Co18}).\par
Our next task is to provide additional insight on the structural financial and economic attributes that differentiate these two states for all experiments. Based on the similarities between the short and long run BTC time frames and the poor convergence of the algorithm for the ETH series, we focus on the long-run BTC series.

\subsection{Hidden States: Financial Properties (BTC 2014-2019)}\label{sec:comparisons}
The results of both the descriptive statistics and the relevant statistical tests are summarized in \Cref{tab:subchains}. Each entry -- BTC price, log-price and log-return series -- consists of two rows that correspond to the subseries of state 1 (upper row) and state 2 (lower row), respectively. The first two columns of \Cref{tab:subchains} verify that the estimated hidden process segments the series into two subseries with high/low mean and variance values for all the examined data series. Log-returns exhibit increased kurtosis in comparison to the initial estimates, cf. \Cref{tab:tests_descriptive}, for both subseries (in particular for state 2). Similarly, the skewness of both subseries has increased and has turned positive with the skewness of the second subseries being again much higher than that of the first (cf. \cite{Tak18}). These distributional properties lead to rejection of normality for either subseries and suggest the presence of heavy-tailed data (phenomena in which exreme events are likely, \cite{Zh18})\footnote{Inclusion of a third hidden state could potentially lead to smoothing of these measurements, cf. \Cref{sec:discussion}.}.\par
The identification of two subchains with different kurtosis and skewness can be a useful tool to investors (\cite{Jo03,Ko93,Di02}). As risk measures, kurtosis and skewness cause major changes to the construction of the optimal portofolio (\cite{Ch97,Con13}), especially in emerging and highly volatile markets (\cite{Ca07}).
%Kurtosis and skewness can be used as a risk measure since they measure the extremes rather than focusing on the mean. \cite{Ch97} notes that the incorporation of skewness into the investor's portfolio decision causes major change in the construction of the optimal portfolio, whereas \cite{Con13} conclude that investors seem to prefer positive skewness and large kurtosis. Additionally, in \cite{Ca07}, the authors  highlight the importance of the incorporation of skewness into the investor's portfolio selection in emerging market industries.
% The reason is that investors would be willing to pay a premium for assets whose return distribution are concentrated on the positive side.
\begin{table}[!htb]
\vspace{-0.1cm}
\centering
\setlength{\tabcolsep}{7pt}
\renewcommand{\arraystretch}{1.3}
\resizebox{\textwidth}{!}{%
\begin{tabular}{llrrrr@{\hskip 18pt}rrrrr}
\clrtwo
&&\multicolumn{4}{c}{\b Descriptive statistics}&\multicolumn{5}{c}{\b Tests}\\
&&Mean&Variance&Kurtosis&Skewness&DF &LBQ& KPSS&VR&JB\\[0.1cm]
\bottomline\\[-0.4cm]
\clrone
\b BTC&Price&4920&$2.1\times 10^7$&2.79&0.81&0.58&0&0.01&0.86&0.00\\
\clrone
&&2133&$7.8\times 10^6$&3.99&1.48&0.79&0.00&0.01&0.14&0.00\\
&Log-Price&7.83&1.84\ph&1.71&-0.41&0.95&0&0.01&0.86&0.00\\
&&6.87&1.49\ph&1.98&0.68&0.97&0.00&0.01&0.45&0.00\\\clrone
&Log-Return&0.0039&0.0050\ph&7.85&0.76&0.00&0.77&0.02&0&0.00\\
\clrone
&&0.0018&0.0023\ph&45.48&2.68&0.00&0.00&0.10&0.08&0.00\\\\[-0.4cm]
%\bottomline\\[-0.4cm]
%\clrone
%\b ETH&Price&320&$7.2\times 10^4$&4.69&1.41&0.34&0.00&0.01&0.18&0.00\\
%\clrone &&282&$3.6\times 10^4$&3.55&1.01&0.27&0.00&0.01&0.99&0.00\\
%&Log-Price&5.33&1.12\ph&4.45&-1.14&0.91&0.00&0.01&0.69&0.00\\
%&&5.35&0.79\ph&5.02&-1.23&0.89&0&0.01&0.27&0.00\\
%\clrone
%&Log-Returns&0.0041&0.0047\ph&5.12&0.30&0.00&0.42&0.02&0.00&0.00\\
%\clrone
%&&0.0132&0.0162\ph&12.98&1.92&0&0.15&0.01&0.00&0.00\\\\[-0.4cm]
\bottomline
\end{tabular}}
\vs
\caption{Descriptive statistics (left panels) and p-values for the time series statistical tests (right panels) for the two (2) BTC price, log-price and log-return \textbf{subseries} -- first and second line of each entry -- which correspond to the two hidden states that were identified by the NHPG model for the whole 1/2014-8/2019 time period.}
\label{tab:subchains}\vspace{-0.8cm}
\end{table}
The asymmetry on the distributions and the difference of volatility between the two subchains can be related to the activity of informed or fundamental vs uninformed, noise or non-fundamental investors (or traders). Intuitively, the activity of uninformed investors leads to periods with higher volatility (cf. \cite{Bau18} and references therein). This is true for state 1 and refines the findings of \cite{Zar19b,Bau18} who attribute the informational inefficiency of BTC not only to its endogenous factors of an emerging, non-mature market but also to the non-existence of fundamental traders. \par
The differences between the two states are further explained by the statistical tests. While the p-values of the Dickey-Fuller (DF) and Jarque-Bera (JB) remain the same as for the combined data series, cf. \Cref{tab:tests_descriptive}, the results for the Ljung-Box-Q (LBQ), KPSS and Variance Ratio (VR) tests unveil different characteristics of the two subseries. In state 2 of the log-return series, the zero hypothesis is rejected for the LBQ test but not for the KPSS and VR tests. This suggests that the subseries defined by state 2 is a random walk with trend stationarity and long memory. These findings are related to (and to some extent refine) the results of \cite{Ji18,La18,Khu18,Me19,Zar19} by determining periods with (state 2) and without (state 1) permanent effects (long memory). The subchain of state 1 stills exhibits richer structure which can be potentially attributed to the combined activity and herding behavior of the non-fundamental traders (\cite{Bo19b,Sil19,St19}).

\subsection{Significant Explanatory Variables: Bitcoin and Ether}\label{sec:covariates}
\begin{figure}[!htb]
\centering
\includegraphics[width=\linewidth, trim=2cm 1cm 2cm 3cm]{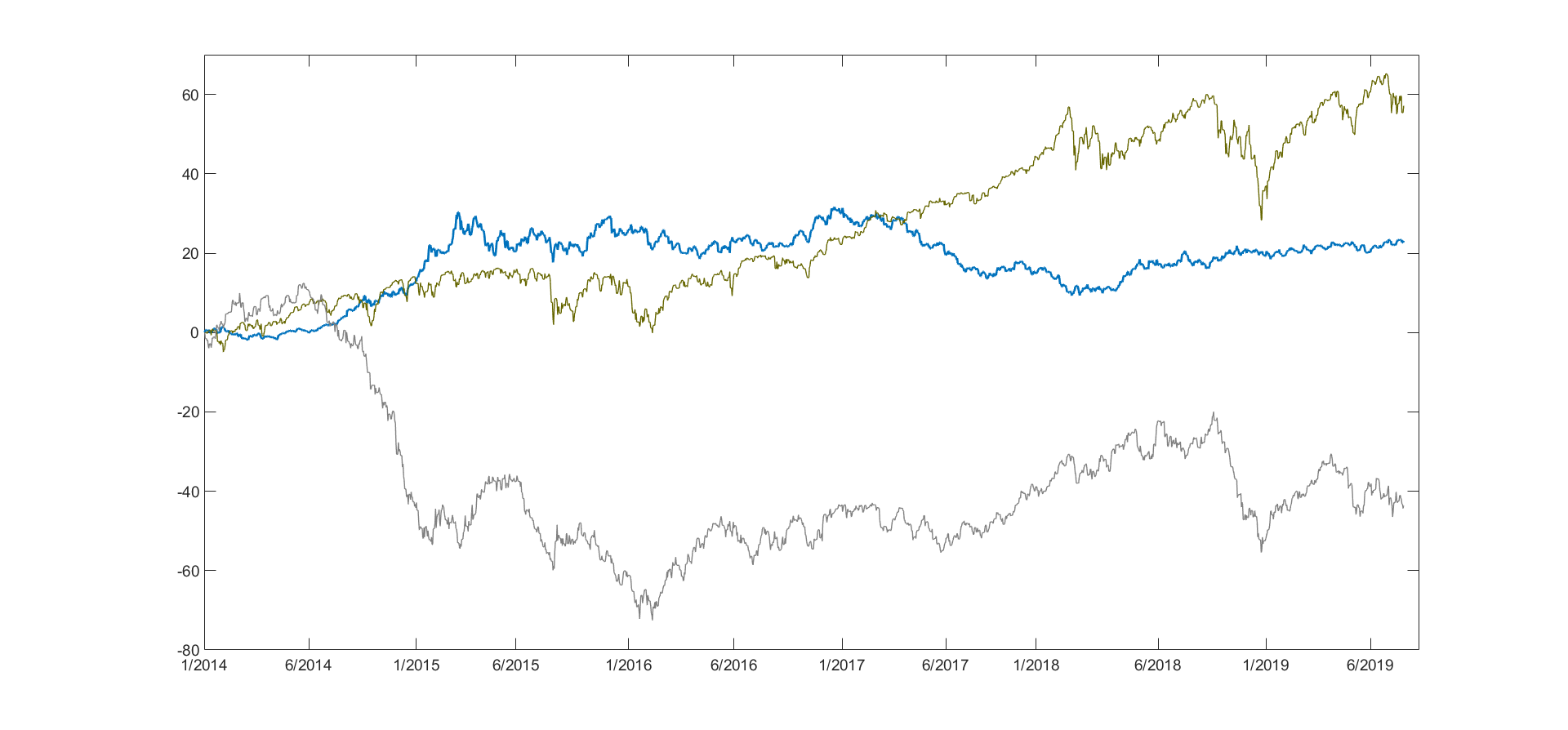}
\caption{Comparison of the USD/EUR exchange rate (blue line), S\&P500 (green line) and Crude Oil Future Prices (gray line) as a percentage of price changes from the initial period. The USD/EUR exchange rate is less volatile than the other predictors.}
\label{fig:predictors}
\end{figure}

The second functionality of the NHPG model is to identify the significant explanatory variables from the set of available predictors that affect the underlying series both linearly, i.e., in the mean equation (observable process), and non-linearly, i.e., in the non-stationary transition probabilities (unobservable process). The algorithm also distinguishes between the variables that are significant in each state. The corresponding results for the BTC log-return series over both the 2014-2019 and 2017-2019 time periods are given in \Cref{tab:postbtc} and the results for the ETH log-return series over the 2017-2019 time period are given in \Cref{tab:posteth}. We use $B_i$ to denote the posterior mean equation coefficients and $\beta_i$ the posterior mean logistic regression coefficients for states $i=1,2$, as described in \Cref{Methodology}.
%\par While the NHPG model has good forecasting ability in financial data, see \cite{Me11,Ko19}, the current results indicate that its forecasting performance with the selected set of predictors, cf. \Cref{tab:variables}, on the BTC and ETH data series, is moderate to poor \stef{We don't see the forecasting performance of the model somewhere nor the in-sample analysis of the next sentence. Is it worth to add an in-sample and a bad forecasting figure?}.\tictoc{Let me check on this. I will need 1-2 days to run again the programs. I removed the out-of-sample period because we rejected the forecasting idea.} However, the in-sample analysis gives valuable insights on the relationship of the cryptocurrency assets and their relation to traditional financial markets. \stef{Same. This paragragh should be either (1) removed, (2) rephrased to fit in the text, (3) remain so if we add some graphs with in-sample and bad forecast. Also, we may want to add similar figures for ETH and small BTC in the appendix? I am thinking that a rich appendix may increase the chances of acceptance.}
\begin{table}[!htb]
\centering
\setlength{\tabcolsep}{8pt}
\renewcommand{\arraystretch}{1.3}
%\resizebox{\textwidth}{!}{%
\begin{tabular}{lrrrr@{\hskip 20pt}rrrr}
\clrtwo
&\multicolumn{8}{c}{\b Estimations BTC}\\
&\multicolumn{4}{c}{\b 2014-2019} & \multicolumn{4}{c}{\b 2017-2019}\\
\b Variables&$B_1$\ph&$B_2$\ph&$\beta_1$\ph&$\beta_2$\ph &$B_1$\ph&$B_2$\ph&$\beta_1$\ph&$\beta_2$\ph\\[0.02cm]
\bottomline\\[-0.4cm]

\clrone USD/EUR&0.00\ph&0.00\ph& $0.19$\ph&$\mb{1.82^{\ast}}$
&-0.01\ph&0.00\ph& $0.48$\ph&$\mb{3.97^{\ast}}$\\
\clrone USD/GBP&$\mb{0.02^{\ast}}$& $\approx 0$\ph&-$\mb{1.35^{\ast}}$& -$\mb{1.68^{\ast}}$
&-0.01\ph&0.00\ph& -1.82\ph&$\mb{4.34^{\ast}}$\\
\clrone USD/JPY&0.00\ph&0.00\ph&0.52\ph&-0.77\ph
&$0.00$\ph&-0.00\ph& -0.53\ph&-0.77\ph\\
\clrone USD/CNY&-0.01\ph&0.00\ph&0.90\ph&0.57\ph
&$\approx0$\ph&0.00\ph& $1.98$\ph&$1.65$\ph\\[0.1cm]

SP500& $\mb{0.04^{\ast}}$ &-0.01\ph&3.90\ph&-1.87\ph
&$0.04$\ph&0.01\ph& $\mb{8.62^{\ast}}$&$1.23$\ph\\
NASDAQ& -0.04\ph&0.00\ph& -1.65\ph&-2.24\ph
&-0.00\ph&-0.02\ph& -$\mb{8.26^{\ast}}$&-2.04\ph\\[0.1cm]

\clrone Silver&0.00\ph&$\approx 0$\ph& 0.22\ph&0.57\ph
&$0.01$\ph&-0.00\ph& -0.42\ph&$1.15$\ph\\
\clrone Gold&0.00\ph&$\approx 0$\ph&$\mb{1.40^{\ast}}$&-0.18\ph
&0.00\ph&0.00\ph& $\mb{1.19^{\ast}}$&-0.35\ph\\
\clrone Oil&$\approx 0$\ph&$\approx 0$\ph&-0.14\ph&$0.00$\ph
&$0.00$\ph&0.00\ph& $\mb{1.27^{\ast}}$&$\mb{1.98^{\ast}}$\\[0.1cm]

VIX& $\approx 0$\ph&$\approx 0$\ph&$0.47$\ph&-0.18\ph
&$\approx0$\ph&$\approx0$\ph& $0.53$\ph&$0.18$\ph\\

EUI&$\approx 0$\ph&$\approx 0$\ph&0.00\ph&-0.00\ph
&$\approx0$\ph&$\approx0$\ph& 0.00\ph&$0.00$\ph\\[0.1cm]

\clrone Blocks&0.00\ph&$\approx 0$\ph&-0.26\ph&-0.07\ph
&-0.01\ph&-0.00\ph& -0.40\ph&$0.06$\ph\\
\clrone Hash&$\mb{0.01^{\ast}}$&0.00\ph&-$\mb{1.84^{\ast}}$&-0.03\ph
&$\mb{0.01^{\ast}}$&$0.01$\ph& -$\mb{2.48^{\ast}}$&$1.60$\ph\\
\clrone Tx-Units&$\approx 0$\ph&0.00\ph&$0.51$\ph&-0.13\ph
&$\approx0$\ph&0.00\ph& $0.53$\ph&-0.50\ph\\\\[-0.4cm]
\bottomline
\end{tabular}
%}
\vs
\caption{Posterior mean estimations for the BTC log-return series in the 2014-2019 (left) and 2017-2019 (right) time periods. $B_1,B_2$ are the mean equation coefficients and $\beta_1,\beta_2$ are the logistic regression coefficients for states 1,2. Statistically significant coefficients (at the 0.05 level) are marked with $\ast$.}
\label{tab:postbtc}\vspace{-0.5cm}
\end{table}
The predictors that have been found significant at the 0.05 level are marked with bold font and $\ast$. The main findings are the following:
\begin{description}[leftmargin=*]
\item[BTC: state 1 vs state 2.] The significant predictors (covariates) that dominate both the observable and the unobservable processes in the more volatile state 1 (cf. \Cref{sec:comparisons}), correspond to more volatile financial instruments such as stock markets (S\&P500 and NASDAQ). By contrast, state 2 is mostly influenced by the more stable exchange rates, cf. \Cref{fig:predictors}. These findings suggest increased speculative activity in state 1 in comparison to fundamental investors in state 2.
\item[BTC: short vs long run.] While the algorithm has identified essentially the same hidden process for both the short and long run windows, cf. \Cref{fig:plotbtc,fig:btc_small}, the significant predictors that affect both the observable and unobservable processes are remarkably different: more volatile for the short run versus more fundamental (monetary) for the long run. In line with \cite{Hor19}, these findings provide evidence for increased speculative behavior in the short run. They also extend BTC's financial and safen haven properties to more recent windows (\cite{Po19,Ba17,Bo17a}). Additionally, they refine the results of \cite{Cor18} and \cite{Cha19} who argue about the differences in the short and long run BTC markets and the hedging properties of BTC against volatile stock indices in time varying periods, respectively. 
\item[ETH vs BTC: short run.] The lower number of significant predictors in the ETH log-return series reflects the inability of the NHPG model to parse the underlying process, cf. \Cref{fig:ploteth}. This differentiates the ETH from the BTC market and provides evidence that ETH is still at its infancy, evolving independently from established economic indicators and fundamentals. Yet, the main -- and somewhat unexpected -- conclusion is that, despite the evident correlation between the prices of BTC and ETH (Pearsons serial correlation 0.62), the two cryptocurrencies are affected by different fundamental financial and macroeconomic indicators over the same time period. 
\end{description}

Finally, an observation that applies to all series is that the current set of predictors cannot fully explain the data volatility. Excluding the miners' activity (as expressed by the Hash Rate) which appears significant in state 1 for all series (both for the observable and the unobservable processes), this observation follows from the small values of the predictors in the mean equation of state 1 (cf. columns $B_1$ in \Cref{tab:postbtc}) and the absence of predictors in the mean equation (observable process) of state 2 (cf. columns $B_2$ in \Cref{tab:postbtc}).

\begin{table}[!htb]
\centering
\setlength{\tabcolsep}{7pt}
\renewcommand{\arraystretch}{1.3}
\resizebox{\textwidth}{!}{%
\begin{tabular}{lrrrr@{\hskip 18pt}lrrrr}
\clrtwo
\multicolumn{10}{c}{\b Estimations ETH 2017-2019}\\\\[-0.4cm]
\b Variables&$B_1$\ph&$B_2$\ph&$\beta_1$\ph&$\beta_2$\ph&
\b Variables&$B_1$\ph&$B_2$\ph&$\beta_1$\ph&$\beta_2$\ph
\\[0.02cm]
\bottomline\\[-0.1cm]

\clrone USD/EUR&-0.01\ph&-0.00\ph&-0.36\ph&-0.38\ph&
USD/GBP& 0.01\ph& -0.06\ph&0.21\ph& 0.59\ph\\
\clrone USD/JPY&-0.01\ph&-0.00\ph&-0.69\ph&0.68\ph&
USD/CNY&-0.02\ph&0.01\ph&-0.14\ph&-0.01\ph\\[0.1cm]

SP500& $\mb{0.06^{\ast}}$&-0.00\ph&2.70\ph&0.13\ph&
VIX& 0.01\ph&0.00\ph&-0.19\ph&0.26\ph\\
NASDAQ&-$\mb{0.06^{\ast}}$&0.00\ph& -$\mb{4.11^{\ast}}$&0.40\ph&
EUI&$\approx 0$\ph&$\approx 0$\ph&-0.00\ph&-$\mb{0.71^{\ast}}$\\[0.1cm]

\clrone Silver&-0.01\ph&0.01\ph& -0.60\ph&0.07\ph&
Blocks&-0.01\ph&$\approx 0$\ph&0.36\ph&2.20\ph\\
\clrone Gold&0.00\ph&-0.01\ph&-0.28\ph&0.21\ph&
Hash&$\mb{0.04^{\ast}}$&0.00\ph&-$\mb{0.46^{\ast}}$&-1.41\ph\\
\clrone Oil&-0.01\ph&-0.01\ph&-0.64\ph&-1.92\ph&
Tx-Units&0.00\ph&0.00\ph&1.03\ph&1.16\ph\\\\[-0.4cm]
\bottomline
\end{tabular}
}
\vs
\caption{Posterior mean estimations for the ETH log-return series in the 2017-2019 time period. Statistically significant coefficients (at the 0.05 level) are marked with $\ast$.}
\label{tab:posteth}\vspace{0.1cm}
\end{table}

\section{Discussion: Limitations and Future Work}\label{sec:discussion}

The application of NHHM modeling in cryptocurrency markets comes with its own limitations. From a methodogical perspective, the main concerns stem from the decision rule for each state which is probabilistic and the exogenously given number of hidden states. In the present study, we used the threshold of $0.5$ to decide transitions from state 1 to state 2 and vice versa. However, in the related financial literature, there are many different approaches even with lower thresholds. Moreover, while two hidden states are generally considered the norm in most financial applications, the current results suggest that it may be worth exploring the possibility of a third hidden state. Alternatively, one may define a \emph{gray zone} for time periods in which the algorithm returns probabilities around 0.5 for both states. This will allow for the identification of periods with high uncertainty about the underlying process and hence, will lead to more scarce, yet more uniform (in terms of distributional properties) subseries.\par
% From a methodological perspective, these are mainly related to two issues: the non time-varying estimated parameters and the exogenously given number of hidden states. Concerning the first, recent studies show that time-varying parameters can reproduce the long-memory of the series (\cite{Ny17}). In the present context, this implies that time-varying parameters may be more suitable to directly model cryptocurrency prices (instead of log-returns as in the present study). While this is interesting for future iterations, studying the log-returns is still more reasonable at the current stage as it enables comparisons with the bulk of the existing literature. \par
%Concerning the limitations on the hidden process, there are actually two things that need to be considered: the number of hidden states and the decision rule for each state.% To determine the optimal number of hidden states, one possibility is to iterate the process of adding or removing states until we obtain the desired distributional properties on the subchains, such as normality or more uniform characteristics (lower kurtosis, less outliers etc.). 
From a contextual perspective, the present approach does not account for qualitative attributes of the predictive variables. For instance, it does not measure centralization of the transactions or alleged fake volumes among different exchanges  (\cite{Ga18,Bo19a}). Coupling the present approach with transaction graph analysis, and user metrics to capture potential market manipulation and the purpose of usage such as speculative trading or exchange of goods and services (\cite{Ch15,Bl17,Ba18}) will lead to improved results. Lastly, as more blockchains transition to alternative consensus mechanisms such as Proof of Stake, further iterations of the current model should also include the underlying technology (e.g., staking versus mining) as a determining factor \cite{Vot20}. At the current stage, such a comparative study is not possible from a statistical perspective, since the market capitalization and trading volume of \qt{conventional} Proof of Work cryptocurrencies is still not comparable to that of coins with alternative consensus mechanisms \cite{Oce20}. The long-anticipated transition of the Ethereum blockchain to Proof of Stake consensus may define such an opportunity in the near future \cite{But20}. \par
Along these lines, extensions of the current model may enrich the set of covariates (explanatory variables) to capture technological features and/or advancements of various cyrptocurrencies, refine the NHPG model with potentially three hidden states and finally, couple the statistical/economic findings with transaction graph analysis. The expected outcome is a more detailed understanding of the financial properties of cryptocurrencies and the assembly of a model with improved explanatory and predictive ability for cryptocurrency markets.

\section*{Acknowledgements}
Stefanos Leonardos and Georgios Piliouras were supported in part by the National Research Foundation (NRF), Prime Minister's Office, Singapore, under its National Cybersecurity R\&D Programme (Award No. NRF2016NCR-NCR002-028) and administered by the National Cybersecurity R\&D Directorate. Georgios Piliouras acknowledges SUTD grant SRG ESD 2015 097, MOE AcRF Tier 2 Grant 2016-T2-1-170 and NRF 2018 Fellowship NRF-NRFF2018-07.

\bibliographystyle{splncs04}
\bibliography{cryptopricebib,
../References/Journal.Economics_Letters/economics_letters_bib,
../References/Journal.Finance_Research_Letters/finance_letters_bib,
../References/Journal.International_Review_of_Financial_Analysis/review_financial_bib,
../References/Journal.Physica_A/physica_bib,
../References/Journal.Research_in_International_Business_and_Finance/business_finance_bib,
%../References/Journal.Digital_Finance/digital_finance_bib,
../References/Journal.Miscellaneous/miscellaneous_bib
}
%../References/Conference.Financial_Cryptography/financial_cryptography_bib}

\appendix
\section{Appendix}\label{app:appendix}
\subsection{Data: Descriptive Statistics and Stationarity Tests}

In \Cref{tab:tests_descriptive}, we summarize the descriptive statistics for the BTC and ETH data series, log-prices and the p-values of the necessary preliminary statistical tests that assess the properties of the given data series prior to the application of the NHPG model. In detail:

\subsubsection{Descriptive statistics}
\begin{description}[leftmargin=*]
\item[Mean \& variance:] We report the mean and variance of prices, log-prices and log-returns of BTC and ETH. As expected, all series exhibit very high (to extreme) volatility. 
\item[Kurtosis:] Based on the kurtosis values, the distributions of all series -- except the log-price BTC series -- are leptokurtic, i.e., they exhibit tail data exceeding the tails of the normal distribution (values above 3), indicating the large number of outliers (extreme values).
\item[Skewness:]Additionally, we report the skewness values, as measure of the asymmetry of the data around the sample mean. If skewness is negative, the data are spread out more to the left of the mean and the opposite if skewness is positive. We observe that the price series are highly right skewed, whereas the skewness of the log-returns for both coins are close to $0$, indicating an approximately symmetrical, around the mean, series.
\end{description}
\begin{table}[!htb]
\centering
\setlength{\tabcolsep}{6pt}
\renewcommand{\arraystretch}{1.3}
\resizebox{\textwidth}{!}{%
\begin{tabular}{llrrrr@{\hskip 15pt}rrrrr}
\clrtwo
&\multicolumn{5}{c}{\b Descriptive statistics}&\multicolumn{5}{c}{\b Tests}\\
&&Mean&Variance&Kurtosis&Skewness&DF &LBQ& KPSS&VR&JB\\[0.05cm]
\bottomline\\[-0.4cm]
\clrone
\b BTC&Price& 3057&$1.37\times 10^7$&4.35&1.40& 0.54&0&0.01&0.80&0.00\\
&Log-Price&7.18&1.82\ph&1.56&0.32&0.96&0&0.01&0.54&0.00\\
\clrone
&Log-Return& 0.0012&0.0016\ph&7.78&-0.27&0&0.31&0.03&0&0.00\\\\[-0.5cm]	
\bottomline\\[-0.5cm]
\b ETH&Price&311&$6.41\times 10^4$&4.95&1.43&0.31&0&0.01&0.43&0.00\\
\clrone
&Log-Price&5.34&1.14\ph&4.42&-1.17&0.91&0&0.01&0.90&0.00\\
&Log-Return&0.0032&0.0037\ph&6.14&0.24&0&0.88&0.01&0&0.00\\	
\bottomline
%\arrayrulecolor{LightCyan2}\hline
\end{tabular}
}
\vs
\caption{Descriptive statistics (left panels) and p-values of the time series statistical tests (right panels) for the BTC and ETH price, log-price and log-return series. DF denotes the Dickey-Fuller test, KPSS the Kwiatkowski-Phillips-Schmidt-Shin test, LBQ is the Ljung-Box Q test, VR the variance ratio test and JB the Jarque-Bera test.} \vspace{-1cm}
\label{tab:tests_descriptive}
\end{table}

\subsubsection{Statistical tests}
Stationarity captures the intuitive idea that certain properties of a (data generating) process are unchanging. This means that if the process does not change at all over time, it does not matter which sample portion of observations we use to estimate the parameters of the process, cf. \Cref{sec:hidden_btc,sec:hidden_btceth}.
\begin{description}[leftmargin=*]
\item[DF-ADF:] First, we report the p-values of the Dickey-Fuller (DF) unit root test\footnote{We also performed the Augmented Dickey-Fuller test with drift $c$, which assesses the null hypothesis of a unit root using the model $y_t=c+\phi y_{t-1}+\beta_1\Delta y_{t-1}+\dots+\beta_p\Delta y_{t-p}+\epsilon_t$ where $\Delta y_t=y_t-y_{t-1}$ and lagged operator $p=7$. The results were the same as the DF test. The null hypothesis was rejected only in the log-returns series for both cryptocurrencies.}. This test assesses the null hypothesis of a unit root using the model $y_t=\phi y_{t-1}+\epsilon_t$. The null hypothesis is $H_0: \phi=1$ under the alternative $H_1: \phi<1$. The $H_0$ was rejected only in the log-return series. The existence of the unit root is one of the common causes of non-stationarity. Intuitively, if a series is unit root nonstationary then the impact of the previous shock  $\epsilon_{t-1}$ on the series has a permanent effect on the series.
\item[LBQ:] To test for serial autocorrelation on the long-run, i.e., to detect if the observations are random and independent over time, we used the Ljung-Box-Q (LBQ) test which assesses the presence of autocorrelations ($\rho$) at lags $p$ in one hypothesis, jointly. The null hypothesis of the LBQ test is $H_0: \rho_1=\dots=\rho_p=0$, under every possible alternative. The null hypothesis was not rejected only for the log-return series and for lags up to $p_{BTC}=10$ and $p_{ETH}=6$, for BTC and ETH respectively. However, when $p_{BTC}>10$ and $p_{ETH}>6$ the null hypothesis was rejected, indicating long-memory (persistent) log-return series. 
\item[KPSS:] The next column presents the p-values of the Kwiatkowsi, Phillips Schimdt, Shin (KPSS) test. The KPSS test assesses the null hypothesis that a univariate time series is trend stationary against the alternative that it is a non stationary unit root process. The test uses the structural model:
$y_t=c_t+\delta_t+u_{1t}$, $c_t=c_{t-1}+u_{2t}$ where $\delta_t$ is the trend coefficient at time $t$, $u_{1t}$ is a stationary process and $u_{2t}$ is an independent and identically distributed process with mean $0$ and variance $\sigma^2$. The null hypothesis is that $\sigma^2 = 0$, which implies that the random walk term ($c_t$) is constant and acts as the model intercept. The alternative hypothesis is that $\sigma^2 > 0$, which introduces the unit root in the random walk. Based on the p-values, we reject all the hypothesis of trend stationarity of the series.
\item[VR:] Additionally, we report the p-values of the Variance Ratio (VR) test which assesses the hypothesis of a random walk. The random walk hypothesis provides a mean to test the weak-form efficiency – and hence, non-predictability – of financial markets, and to measure the long run effects of shocks on the path of real output in macroeconomics, see \cite{Ch09} and references therein. The model under the $H_0$ is $y_t = c + y_{t-1} + \epsilon_{t}$, where $c$ is a drift constant and $\epsilon_t$ are uncorrelated innovations with zero mean. The random walk hypothesis is rejected only in the log-return series for both coins. Essentially, the rejection of the random walk hypothesis shows that there exists information that can be used in explaining the movement of the returns.  
\item[JB:] Lastly, we report the Jarque-Bera (JB) test, as a normality test. Based on these results, all the series are not normally distributed.
\end{description}

\end{document}